\documentclass[prl,aps,twocolumn,showpacs,preprintnumbers,amsmath,amssymb,superscriptaddress]{revtex4}
\usepackage{graphicx}
\usepackage{simplewick}
\usepackage{color}
\usepackage[english]{babel} 
\usepackage{hyphenat}
\newcommand{\ket}[1]{\ensuremath{\left|{#1}\right\rangle}}
\newcommand{\bra}[1]{\ensuremath{\left\langle{#1}\right|}}

\begin{document}

\title{The Fermi problem with artificial atoms in circuit QED}

\author{Carlos Sab\'in}
\affiliation{Instituto de F\'{\i}sica Fundamental, CSIC,
  Serrano 113-B, 28006 Madrid, Spain}
\email{csl@iff.csic.es}

\author{Marco del Rey}
\affiliation{Instituto de F\'{\i}sica Fundamental, CSIC,
  Serrano 113-B, 28006 Madrid, Spain}
\email{csl@iff.csic.es}

\author{Juan Jos\'e Garc\'ia-Ripoll}
\affiliation{Instituto de F\'{\i}sica Fundamental, CSIC,
  Serrano 113-B, 28006 Madrid, Spain}
\email{csl@iff.csic.es}

\author{Juan Le\'on}
\affiliation{Instituto de F\'{\i}sica Fundamental, CSIC,
  Serrano 113-B, 28006 Madrid, Spain}

\begin{abstract}
We propose  a feasible experimental test  of a 1-D version of the Fermi problem using superconducting qubits.  We give an explicit non-perturbative proof of strict causality in this model, showing that the probability of excitation of a two-level artificial atom with a dipolar coupling to a quantum field is completely independent of the other qubit until signals from it may arrive. We explain why this is in perfect agreement with the existence of nonlocal correlations and previous results which were used to claim apparent  causality problems for  Fermi's two-atom system. 
\end{abstract}


\maketitle

Information cannot travel faster than light. But in quantum theory, correlations may be established between spacelike separated events. These facts are not contradictory, since correlations need to be assisted with classical communication in order to transmit information.

 The two physical phenomena above arise in a natural fashion in the following situation,  which is the so-called Fermi problem \cite{fermi}, originally proposed by Fermi to check causality at a microscopic level. At $t=0$ a two-level neutral atom $A$ is in its excited state and a two-level neutral atom $B$ in its ground state, with no photons present. If $A$ and $B$ are separated by a distance $r$ and $v$ is the speed of light, can $A$ excite $B$ at times $t<r/v$?  Fermi 's answer was negative but his argument had a mathematical flaw. When a proper analysis is carried on, fundamental quantum theory questions arise due to the interplay between causal signaling and quantum non-local phenomena. 
 
These issues led to a controversy \cite{hegerfeldtfer,yngvasonfer,yngvasontipo3,powerthiru} on the causal behavior of the excitation probability of qubit B, whose conclusions were never put to experimental test. A notorious claim on causality problems in Fermi's two-atom system was given in \cite{hegerfeldtfer}. The reply of \cite{yngvasonfer} was in the abstract language of algebraic field theory and the proof of strict causality of \cite{powerthiru} is perturbative,  although given to all orders in perturbation theory.  The Fermi problem is usually regarded just as a gedanken experiment, and remains untested, essentially because interactions between real atoms cannot be switched on and off.

With this work we give a complete description of the problem in a physical framework in which predictions can be verified. This framework will be circuit QED  which  can be regarded as a 1-D version of Quantum Electrodynamics (QED) with two-level (artificial) atoms, a testbed which makes it possible to control the interaction and tune the physical parameters. We complete previous descriptions made of the problem and explain how there are no real causality issues for Fermi's two-atom system. We give an explicit non-perturbative proof of strict causality in these setups, showing that the probability of excitation of qubit $B$ is completely independent of qubit $A$ for times $t<r/v$ and for arbitrary initial states. As a matter of fact, this comes as a manifestation of the nonsignaling character of the quantum theory~\cite{gisin}. We also show how this is compatible with the existence of nonlocal correlations at times $0<t<r/v,$ a fact pointed out in various theoretical proposals to entangle qubits at arbitrarily short times~\cite{reznik,franson,conjuan,conjuanjo}. More precisely, we give a non-perturbative proof of the fact that the probability of $B$ being excited and $A$ in the ground state is finite and $r$-dependent  at any time, even for $t<r/v$. We provide also a physical and intuitive explanation of why the conclusions in \cite{hegerfeldtfer}, even if mathematically sound, do not apply to the causality problem. At the end we discuss the time dependence predicted in our model for the various excitation probabilities and suggest a feasible experimental test of causality using superconducting circuits.

 In what follows  we focus
on a practical setup of circuit-QED, with two qubits, $A$ and $B,$
interacting via a quantum field. The qubits have two stationary states
$\ket{e}$ and $\ket{g}$ separated by an energy $\hbar\Omega$ and
interact with a one-dimensional field, $V(x),$ which propagates
along a one-dimensional microwave guide that connects them
\begin{eqnarray}
  V(x)=i\,\int_{-\infty}^{\infty}dk\, \sqrt{N\omega_k}\,e^{ikx}a_k +\mathrm{H.c.} \label{a}
\label{field}
\end{eqnarray}
This field has a continuum of Fock operators $[a_k,a^{\dag}_{k'}]=\delta(k-k'),$ and a linear spectrum, $\omega_k =
v|k|$, where $v$ is the propagation velocity of the field. The normalization and the speed of photons, $v=(cl)^{-1/2},$ depend on the microscopic details such as the capacitance and inductance per unit length, $c$ and $l.$ We will assume qubits that are much smaller than the relevant wavelengths, $\lambda=v/\Omega,$ and are well separated. Under these conditions the Hamiltonian, $H = H_0 + H_I,$  splits into a free part for the qubits and the field 
\begin{equation}
  H_0 = \frac{1}{2}\hbar\Omega(\sigma^z_A + \sigma^z_B) + \int_{-\infty}^{\infty}dk\, \hbar\omega_k
  a^{\dagger}_ka_k \label{b}
\end{equation}
and a point-like interaction between them
\begin{equation}
  H_I = \sum_{J=A,B} d_J\,\sigma_J^x\,V(x_J) \label{c}
\end{equation}
Here $x_A$ and $x_B$ are the fixed positions of the atoms, and $d_J\, \sigma^x_J$ is equivalent to the dipole moment in the case of atoms interacting with the electromagnetic field.

The original formulation of the Fermi problem begins with an initial state
\begin{equation}
 \ket{in} = \ket{e_A\,g_B\,0} \label{d}
\end{equation}
in which only qubit $A$ has been excited, while $B$ and the field remain in their ground and vacuum states, respectively. The total probability of excitation of qubit $J$ is the expectation value of the projector onto the excited state $\mathcal{P}^e_J=\ket{e_J}\bra{e_J}.$ In the Heisenberg picture
\begin{equation}
P_{eJ}=\bra{in}\mathcal{P}^e_J (t)\ket{in},\quad J\in\{A,B\}.
\label{eq:probability}
\end{equation}
We will prove that for $vt<r$ the probability $P_{eB}$ is \textit{completely independent of the state of qubit $A$ for all initial states.} In the Heisenberg picture this amounts to showing that there appears no observable of $A$ in the projector $\mathcal{P}^e_B(t)$ for $vt<r.$ Our proof begins by solving formally the Heisenberg equations for $\mathcal{P}^e_J$
\begin{equation}
\mathcal{P}^{e}_J (t)- \mathcal{P}^{e}_J (0)=- \frac{d_J}{\hbar} \int_0^t dt'\sigma^y_J (t') V(x_J,t').
\label{eq:evolution}
\end{equation}
Integrating the Heisenberg equations of the operators $a_k$ and $a_k^{\dagger}$ and inserting them in Eq.~(\ref{a}), the total field evaluated at $x$ in Heisenberg picture is decomposed
\begin{equation}
V(x,t)= V_0(x,t)+V_A(x,t)+V_B(x, t)
\label{eq:fielddecomposed} 
\end{equation}
into the homogenous part of the field
\begin{equation}
V_0(x,t)=i\,\int_{-\infty}^{\infty}dk\,\sqrt{N\omega_k}\,e^{i(kx-\omega t)}a_k +\mathrm{H.c.}
\label{eq:fielddecomposed1} 
\end{equation}
and the back-action of $A$ and $B$ onto the field
\begin{eqnarray}
&&V_J(x,t)=\frac{-id_J\,N}{\hbar}\times
\label{eq:fieldgeneral}\\
{}&&\quad\times \int_0^t \sigma_J^x (t')
\int_{-\infty}^{\infty}\omega_ke^{ik(x-x_J)-i\omega_k\,(t-t')}dkdt'
+\mathrm{H.c.}\nonumber
\end{eqnarray}
Eqs.~(\ref{eq:evolution}) translates into a similar decomposition for the  probabilty $\mathcal{P}^e_B$ with three terms
\begin{equation}
\mathcal{P}^e_B (t)= \mathcal{P}^{e}_{B0} (t)+ \mathcal{P}^{e}_{BB} (t)+ \mathcal{P}^{e}_{BA} (t)
\label{eq:causality}
\end{equation}
which are proportional to $V_0, V_B$ and $V_A,$ respectively.  The only explicit dependence on $A$ may come from $\mathcal{P}^e_{BA}$ through $V_A(x_B,t).$ Manipulating Eq.~(\ref{eq:fieldgeneral}) gives
\begin{eqnarray}
V_A (x_B,t)=\frac{-2\pi d_A N}{\hbar} \frac{d}{dr}\left[\sigma_A^x\left(t-\frac{r}{v}\right)\,\theta\left(t-\frac{r}{v}\right)\right]\label{eq:field2}
\end{eqnarray}
where the Heaviside function $\theta$ shows that strictly $\mathcal{P}^e_{BA}(x_B,t)=0$ for $vt<r,$ and no such dependence is possible. We still have to analyze a possible implicit dependence on A through $\mathcal{P}^e_{BB},$ whose expression is
\begin{eqnarray}
\mathcal{P}^e_{BB} (t)&=&\frac{id_B^2N}{\hbar^2}\int_0^t dt'\int_0^{t'}dt''\sigma_B^y (t')\sigma_B^x (t'')\nonumber \\ & &\int_{-\infty}^{\infty}dk\,\omega_k e^{-i\,\omega_k(t'-t'')}+ \mathrm{H.c.} \label{eq:implicit}
\end{eqnarray}
The only implicit dependence could come through the evolution of  $\sigma_B^{x,y} (t),$ but again this is not the case. Since $[\sigma_B^x,H_I]=0$, the evolution of $\sigma_B^x$ does not involve the field in any way, and for $\sigma_B^y (t)$ we have that $\dot{\sigma}_B^y (t)=\Omega\,\sigma_B^x(t)/2\, - \frac{d_B}{\hbar} V(x_B,t)\sigma_B^z (t)$ so using again Eq. (\ref{eq:fielddecomposed}) and Eq. (\ref{eq:field2}) we see that  the A-dependent part of $P^e_{BB}$ is 0 for $vt<r$.  Thus $\mathcal{P}^e_B$ may be finite but is completely independent of qubit $A$ for $vt<r$, as we wanted to show.

So far, we have demonstrated that although $P^e_B (t)$ is non-zero for $vt<r,$ the only non-zero contribution is $\mathcal{P}^e_{B0},$ which is not sensitive to the qubit $A$ and thus cannot be used to transmit information between the qubits. Now we will show that this result is compatible with the existence of correlations for $vt<r$. For instance, we consider the probability of finding qubit $B$ excited and qubit $A$ on the ground state $P_{eB,gA}$, which is:
\begin{equation}  
P_{eB,gA}=\bra{in}\mathcal{P}^e_B (t)\mathcal{P}^g_A (t)\ket{in},
\label{eq:correlation}
\end{equation}
where $\mathcal{P}^g_A=\openone-\mathcal{P}^e_A$. Using Eqs.~(\ref{eq:evolution}), (\ref{eq:fielddecomposed}) and (\ref{eq:causality}), we find a term in this probability which is proportional to $\mathcal{P}^{e}_{BB} \, \mathcal{P}^{g}_{AA}$ and thus to  $V_B(x_B,t)\,V_A(x_A,t)$. From Eq.~(\ref{eq:fieldgeneral}) we obtain: $V_J(x_J,t)\propto\frac{d}{dt}\{\sigma_J^x(t) \theta(t)$\}. Therefore, we conclude that in  (\ref{eq:correlation}) there is an unavoidable dependence on $A$ at any $t>0,$ but this is not a causality violation because correlations alone cannot transmit information.

At this point it remains a single question:  How can the A-dependent part of  $P_{eB}$ be zero while the one of  $P_{eB,gA}$  is nonzero for $vt<r$? To better understand it we need less formal results that rely on perturbative expansions, but we would like to remark here that the conclusions above are valid to all orders in perturbation theory.
\begin{figure}[t]
\centering
\includegraphics[width=\linewidth]{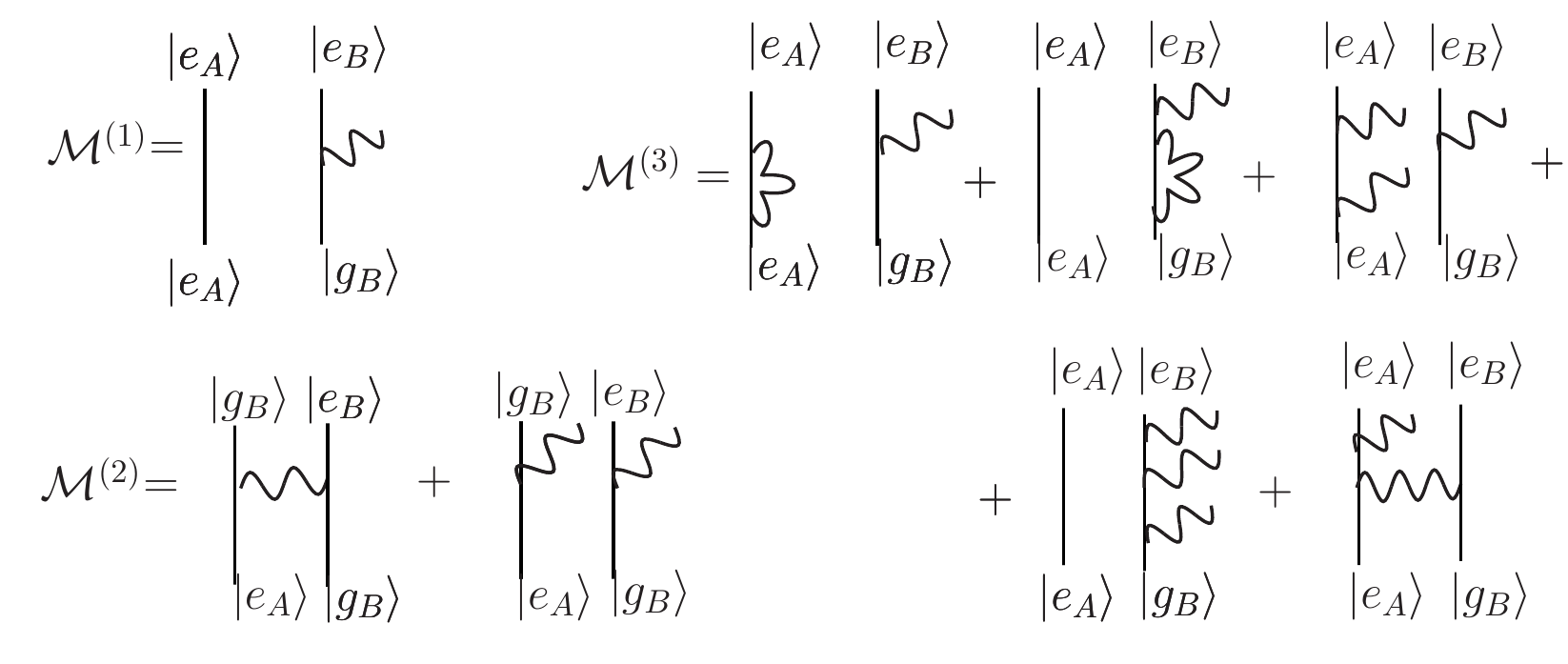}
\caption{The terms of order $d_J, d_J^2$ and $d_J^3$ contributing to the amplitude for exciting qubit $B$. } \label{Fig.1}
\end{figure}
To obtain the total probability of excitation of qubit $B$ $P_{eB}$ to a given order in perturbation theory, one has to expand to a certain order the operators appearing in Eqs.~(\ref{eq:evolution}), (\ref{eq:fielddecomposed}), (\ref{eq:fieldgeneral}). The different terms in the expansion can be related to the probabilities of the different physical processes involved. Fig.~\ref{Fig.1} shows the diagrams of the different amplitudes contributing to $P_{eB}$ up to the fourth order in $d_J.$ The lowest order amplitude contributing to a final excited $B$ qubit is of order $d_J,$ which means that terms up to order $d_J^3$ have to be considered. The only terms leading to this final state will be $\mathcal{M}^{(1)} = v_B,\;\, \mathcal{M}^{(2)} = x + u_A\,v_B,\,\;  \mathcal{M}^{(3)}= a' v_B +u_Av_Av_B+ v_Bu_Bv_B+\mathcal{\delta M}^{(3)}$
where $u_{J}$ ($v_J$) represent the amplitude for single photon emission at qubit $J$ when the qubit is initially in the ground (excited) state, $x$ is the amplitude for photon exchange, $a_J$  are the  radiative corrections of qubit $J$, and finally  $\mathcal{\delta M}^{(3)}$ is the amplitude for photon exchange accompanied by a single photon emission at qubit $A$. Notice that some of these processes are only possible beyond the rotating wave approximation, which breaks down for strongly coupled circuit-QED setups~\cite{ultrastrong2} as the ones considered later. Keeping only terms up to fourth order, we have for the probability to get B excited at a time $t$
\begin{eqnarray}
P_{eB}(t) &=& |\mathcal{M}^{(1)}|^2 +|\mathcal{M}^{(2)}|^2 + 2\, Re\{\mathcal{M}^{(1)^*} \mathcal{M}^{(2)}\} \nonumber\\ 
&+&2\, Re\{\mathcal{M}^{(1)^*} \mathcal{M}^{(3)}\}+ \mathcal{O}(d^5)\label{s}
\end{eqnarray}
The final states in $\mathcal{M}^{(1)}$ are orthogonal to those in $\mathcal{M}^{(2)}$ and to the three photon terms in $\mathcal{M}^{(3)}$. Hence, their interference vanishes. Besides, we are only interested in the $A$-dependent part of the probability, so we can  remove the $r$-independent terms left in (\ref{s}), marking the remaining contributions with a superscript ${}^{(r)}$
\begin{equation}
\label{eq:perturbation}
P_{eB}^{(r)}(t) = |\mathcal{M}^{(2)}|^{2^{(r)}} + 2 Re\{\mathcal{M}^{(1)^*}\mathcal{\delta M}^{(3)}\} + \mathcal{O}(d^5).
\end{equation}
The first term actually gives $P_{eB,gA}$ up to the fourth order, it is positive and $A$-dependent at all times, as shown in Fig.~\ref{fig2}a. The second term is not a projector onto any physical state, but an interference term which has the effect of canceling out exactly the first term for $vt<r$ but not for $vt>r$ (cf. Fig.~\ref{fig2}b). In a nutshell, interference seems to be the physical mechanism that operates at all orders in perturbation theory to give the causal behavior of the total probability of excitation that we had previously shown.

These perturbative results cast new light on the controversy on the Fermi problem and help us understand \textit{why} our results do not contradict those of Hegerfeldt~\cite{hegerfeldtfer}. Hegerfeldt proved mathematically that the expectation value of an operator consisting of a sum of projectors cannot be zero for all the times $vt<r,$ unless it is zero at any time. Indeed, the expectation value of $\mathcal{P}^e_B (t)$ cannot be zero for all $vt<r,$ for it always contains the contribution $\mathcal{P}^e_{B0}$ from Eq.~(\ref{eq:fielddecomposed1}).  However, as we showed non-perturbatively, the actual relevant question for causality is whether the expectation value of $\mathcal{P}^{e}_{BA}(t)$ vanishes for $vt<r$ or not, since only this part of the probability is sensitive to qubit $A$ and could be used to transmit information. Besides, according to our above perturbative results to fourth order, the $r$-dependent part of the probability, that is the expectation value of  $\mathcal{P}^{e}_{BA}(t)$, is not a mere sum of projectors, but also contains interfering terms. Thus, Hegerfeldt's result does not apply  and  $\mathcal{P}^{e}_{BA}(t)$ can be zero for $vt<r$ as is actually the case. Both results are in accord with a general fact of Relativistic Quantum Field Theory: two global states can not be distinguished locally with the aid of a local projector annihilating one of the states, since the local observable algebras are Type III von Neumann algebras (See \cite{yngvasonfer,yngvasontipo3} for a discussion).
\begin{figure}[t]
\begin{center}
\includegraphics[width=0.8\linewidth]{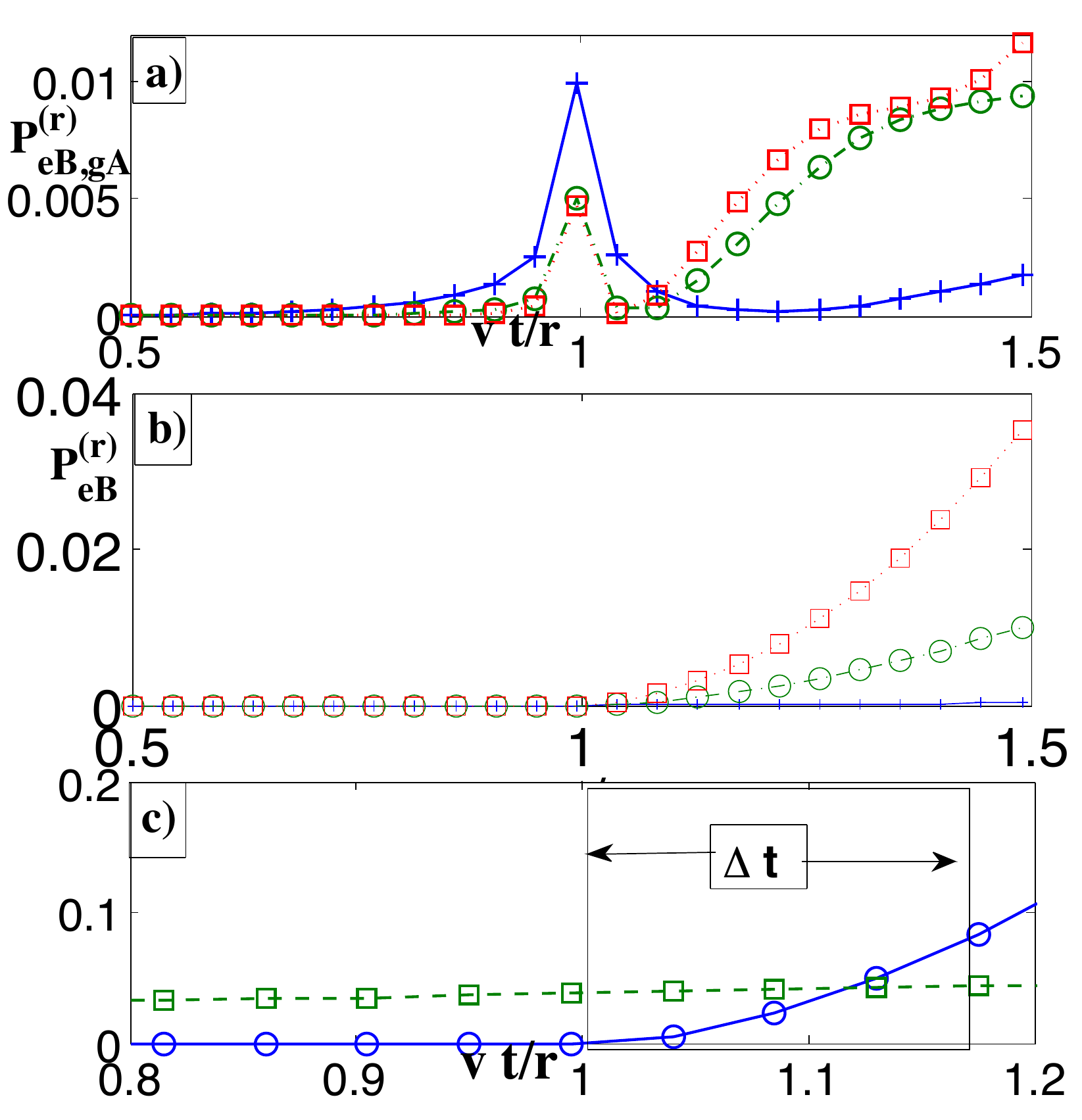}
\end{center}
\caption{(a) $P^{(r)}_{eB,gA}$ and (b) $P^{(r)}_{eB}$ versus $v t/r$ for $\Omega r/v= \frac{\pi}{2}$ (blue, crosses), $\pi$ (red, squares), and $2\pi$ (green, circles) with $K_{A,B}=0.0225$. For $vt<r$ the qubits are spacelike separated, but there are correlations between them and figure (b) shows the expected causal behavior. (c) $P^{(r)}_{eB}$  (blue, circles) and  $|\mathcal{M}^{(1)}|^2$ (green, squares) vs.  $v t/r$ for $K_A=0.20$, $K_B=0.04$ and a separation of one wavelength $r = 2\pi v/\Omega_{A,B}.$ With this data and $\Omega/2\pi\simeq1$GHz we have $\Delta\,t\simeq1$ns. (Color online) } \label{fig2}
\end{figure}
 We will now suggest an experiment to check the causal behavior of  $P_{eB}.$ For this we need to control the interaction time at will to access the regions at both sides of $t=r/v.$ This, which is highly unrealistic with real atoms, becomes feasible in circuit-QED. While the ideas are valid for both inductive and capacitive couplings, we will focus on using a pair of three-junction flux qubits \cite{mooij,reviewflux}. Each of the qubits is governed by the Hamiltonian $H_{0J}= \frac{1}{2}\epsilon_J \sigma^z_J+ \frac{1}{2}\Delta_J\sigma^x_J.$
The energy $\epsilon_J=2I_p\delta\Phi_{xJ},$ is approximately linear in the external magnetic flux, $\delta\Phi_{xJ},$ measured from the degeneracy point, and we assume that the gap $\Delta_J$ is fixed. The result is a qubit energy difference $\Omega_J(\delta\Phi_{xJ})=\sqrt{(2\,I_p\,\delta\Phi_{xJ})^2+\Delta_J^2}.$

The coupling between the qubit and the microwave photons is ruled by the dimensionless ratio
\begin{equation} \label{eq:couplingstrength}
K_{J}=\frac{4d_{J}^2N}{\hbar^2 v}= 2\left(g/\Omega_J\right)^2.
\end{equation}
Here $g=d_J\sqrt{N\Omega}/\hbar$ is the coupling strength between a qubit and the cavity that would be obtained by cutting the transmission line to be perfectly resonant with the qubit transition. These numbers enter the qubit excitation probability computed before (\ref{eq:perturbation}) through the product $P^{(r)}_{eB}(t) \propto K_AK_B.$ Since $K_J\propto 1/\Omega_J,$ we may use the external fluxes to move from a weakly coupled regime with no qubit excitations, $\Omega_J \ll g,$ to the maximum coupling strength, $\Omega_J\simeq \Delta_J~(\delta\Phi_{xJ}=0).$

Let us first discuss how to prepare the initial state (\ref{d}) of the Fermi problem. We assume that the system starts in a ground state of the form $\ket{g_A\,g_B\,0}$. This is achieved cooling with a large negative value of $\delta\Phi_{xJ}$ on both qubits, which ensures a small value of $g/\Omega_J$ and $K_J.$ We estimate that couplings  $g/\Omega_J<0.15$ and $\Omega_J \sim 1.5$GHz lower the probability of finding photons in the initial state below $5\times10^{-3},$ both for vacuum and thermal excitations. Both magnetic fluxes are then raised up linearly in time, $\delta\Phi_{xJ}=\alpha_J t,$ to prepare the qubits. Using a Landau-Zener analysis~\cite{zener} of the process we conclude that an adiabatic ramp   $\alpha_B\ll \pi\Delta_B^2 / 4 \hbar I_p$ of qubit $B$ followed by a diabatic ramp \cite{johanssoncasimir1, johanssoncasimir2} $\alpha_A\gg \Delta_A^2/\hbar2I_p$ of qubit A, leads to the desired  state $\ket{e_A\,g_B\,0}$ with a fidelity that can be close to 1, depending only of $ \alpha_A$, $\alpha_B$ as derived from the Landau-Zener formula.  Note that the minimum gap $\Delta_B$ has to be large enough to ensure that the qubit-line coupling of B remains weak and the qubit does not ``dress" the field with photons.

Once we have the initial state, both magnetic fluxes must take a constant value during the desired interaction time.  After that, measurements of the probability of excitation of qubit B can be performed with a pulsed DC-SQUID scheme \cite{measurementqubits1, measurementqubits2}. The timescale of the ``jump''   of the probability around $t=r/v$ for qubit frequencies in the range of GHz and a separation of one wavelength $r=2\pi v/\Omega$ is $\Delta t\simeq 1$ns [Fig.~\ref{fig2}c]. Although the total measurement of the SQUID may take a few $\mu$s, the crucial part is the activation pulse ($\sim 15$ns) in which the SQUID approaches its critical current and may switch depending on the qubit state. During this activation period the SQUID and the qubit are very strongly coupled $(g\sim\mathrm{GHz}),$~\cite{measurementqubits1} and the qubit is effectively frozen. The time resolution of the measurement is thus determined by the ramp time of the activation pulse, which may be below nanoseconds. Among the sources of noise that are expected, the short duration of the experiment, well below $T_1$ and $T_2$ of usual qubits, makes the ambient noise and decoherence pretty much irrelevant. Thermal excitations of the qubits and the line may be strongly suppressed by using larger frequencies ($>1.5$GHz). The most challenging aspect is the low accuracy of SQUID measurements, which are stochastic, have moderate visibilities~\cite{measurementqubits1} and will demand a large and careful statistics.

On the technical side, it is important to choose carefully the coupling regimes. If we wish to compare with perturbation theory, we need $K_J \ll 1.$ However, at the same time the product $K_AK_B$ must take sizable values for $P_{eB}\propto K_AK_B$ to be large. And we need to discriminate the causal signal from the $r$-independent background of the probability of excitation, whose main contribution is $|\mathcal{M}^{(1)}|^2\propto K_B$. Thus, a good strategy would be to work with $K_A>K_B.$  In Fig. \ref{fig2}c we show that it is possible to achieve a regime in which the perturbative approximations are still valid and the $r$-dependent part of $P_{eB}$ is comparable to 
$|\mathcal{M}^{(1)}|^2$ in the spacetime region of interest $v t\simeq r$. 

 To conclude, we have considered a system of two superconducting qubits coupled to a transmission line, which can be suitably described in the framework of 1-D QED with two-level (artificial) atoms. Starting from an initial state with qubit A excited, qubit B in the ground state and no photons, we have illustrated the causal character of the model  showing that the probability of excitation of qubit B is completely independent of qubit A when $vt<r$. We have also shown that this is in agreement with the existence of nonlocal correlations and we have used perturbative computations to see the physical mechanism underlying the causal behavior. Finally, we have suggested an experiment feasible with current technology that would solve the controversy on the Fermi problem.
 
The authors would like to acknowledge S. Pascazio and J. Yngvason for useful discussions. This work is supported by Spanish MICINN Projects FIS2008-05705, FIS2009-10061, and CAM research consortium QUITEMAD S2009-ESP-1594. M. del Rey acknowledges the support from CSIC JAE-PREDOC2010 grant and Residencia de Estudiantes.

\end{document}